
\input phyzzx

\twelvepoint

\def\dmm{{\partial}_=}
\def\dpp{{\partial}_{\ne}}
\def\cX{{\cal X}}

\def\a{\alpha'}
\def\eab{\epsilon_{AB}}
\def\eyz{\epsilon_{YZ}}
\def\eabp{\epsilon_{A'B'}}
\def\eyzp{\epsilon_{Y'Z'}}

\def\dalemb#1#2{{\vbox{\hrule height .#2pt
        \hbox{\vrule width.#2pt height#1pt \kern#1pt
                \vrule width.#2pt}
        \hrule height.#2pt}}}

\REF\Wittena{E. Witten, J. Geom. Phys. {\bf 15} (1995) 215}
\REF\ADHM{M.F. Atiyah, V.G. Drinfeld, N.J.
Hitchen and Y.I. Manin, Phys. Lett. {\bf 65A} (1978) 185}
\REF\WW{R.S. Ward and R.O. Wells Jr, Twistor Geometry and
Field Theory, (Cambridge University Press, Cambridge,
1992)}
\REF\GR{S.J. Gates Jr and L. Rana, Phys. Lett. {\bf
345B} (1995) 233}
\REF\GS{A. Galperin and E. Sokachev,
"Manifest Supersymmetry and the ADHM Construction of
Instantons", JHU-TIPAC-94021, BONN-TH-94-27,
hep-th/9412032}
\REF\GSt{A. Galperin and E. Sokachev,
"Supersymmetric Sigma models and the 't Hooft Instantons",
JHU-TIPAC-94010, BONN-TH-94-08, hep-th/9504124}
\REF\HPT{G.Papadopoulos, C. Hull and P.K. Townsend, Phys.
Lett. {\bf 316B} (1993) 291}
\REF\PT{G.Papadopoulos and P.K. Townsend, Class.
Quan. Grav. {\bf 11} (1994) 515}
\REF\NDL{N.D. Lambert, "Two
Loop Renormalization of Massive (p,q) Supersymmetric Sigma
Models", DAMTP R/94/42, hep-th/9510130}
\REF\Back{P.S. Howe,
G. Papadopoulos and K.S. Stelle, Nucl. Phys.{\bf B296}
(1988) 26}
\REF\HPb{P. Howe and G. Papadopoulos, Nucl.
Phys. {\bf B289} (1987) 264}
\REF\HPa{P. Howe and G.
Papadopoulos, Nucl. Phys. {\bf B381} (1992) 360}
\REF\HPc{P.
Howe and G. Papadopoulos, Class. Quan. Grav. {\bf 5} (1988)
1647}
\REF\CHS{C.G. Callan Jr., J.A. Harvey and A.
Strominger, Nucl. Phys. {\bf B359} (1991) 611}
\REF\Wittenb{E. Witten, "Some Comments on String Dynamics",
to appear in the proceedings of {\it Stings '95}, USC,
March 1995, hep-th/9507121}
\REF\GT{G. Gibbons and P.K.Townsend, Phys. Rev. lett. {\bf
71} (1993) 3754}

\pubnum={DAMTP R/95/43\cr hep-th/9508039}

\titlepage

\title{\bf Quantizing the (0,4) Supersymmetric ADHM Sigma Model}

\centerline{N.D. Lambert\foot{nl10000@damtp.cam.ac.uk}}

\address{D.A.M.T.P., Silver Street\break
         University of Cambridge\break
         Cambridge, CB3 9EW\break
         England}

\vfil

\abstract

We discuss the quantization of the ADHM sigma model. We show that the
only quantum contributions to the effective theory come from the chiral
anomalies and compute the first and second order terms. Finally the limit
of vanishing instanton size is discussed.

\endpage

\chapter{Introduction}

Despite their central role in string compactifications, chiral $(0,q)$
supersymmetric sigma models have received less attention in comparison to
the non chiral models due to difficulties involved with their
construction and quantization. Recently there has been considerable
interest in the use of massive linear sigma models to construct string vacua
as the infrared conformal fixed point of the renormalization group flow. This
method allows one to construct a large class of string vacua including those
with chiral supersymmetry. In particular an interesting paper by Witten
[\Wittena] discusses a class of massive linear sigma models possessing
on-shell (0,4) supersymmetry which flow in the infrared to conformally
invariant sigma models describing ADHM instantons [\ADHM,\WW].

Previous work on the ADHM sigma model
has focused primarily on classical aspects of the $(0,4)$ supersymmetry
multiplet used and in particular the construction of off-shell superfield
formalisms [\GR,\GS,\GSt]. Here we will study the models quantum properties
and it's rich interplay between geometry and field theory in detail. The
general $(p,q)$ supersymmetric massive sigma model has been constructed
before [\HPT,\PT] and it's quantization is discussed to two loop order in
[\NDL]. We will show here that the ADHM sigma model is ultraviolet finite to
all orders of perturbation theory and integrate out the massive fields to
obtain the low energy effective theory. Due to anomalies this theory has
interesting non trivial properties and we obtain the quantum corrections to
order $\a^2$ by requiring that the anomalies are appropriately canceled. We
conclude by making some comments about the case where the instanton size
vanishes.

\chapter{The ADHM Sigma Model}

In [\Wittena] Witten constructs an on-shell (0,4) supersymmetric linear sigma
model which parallels the ADHM construction of instantons [\ADHM]. The model
consists of $4k$ bosons $X^{AY}$, $A=1,2,\ Y=1,2...,2k$ with
right handed superpartners $\psi^{A'Y}_-$, $A'=1,2$. There is also a
similar multiplet of fields $\phi^{A'Y'},\ \chi_-^{AY'}$ $Y'=1,2...,2k'$. In
addition there are $n$ left handed fermions $\lambda^a_+$, $a=1,2...,n$. The
$A,B...$ and $A',B'...$ indices are raised (lowered) by the two
by two antisymmetric tensor $\epsilon^{AB}$ ($\epsilon_{AB}$),
$\epsilon^{A'B'}$ ($\epsilon_{A'B'}$). The $Y,Z...$ and $Y',Z'...$ indices are
raised (lowered) by the invariant tensor of $Sp(k)$, $Sp(k')$ respectively
which are also denoted by $\epsilon^{YZ}$ ($\epsilon_{YZ}$),
$\epsilon^{Y'Z'}$ ($\epsilon_{Y'Z'}$).

The interactions are provided for by a tensor $C^a_{AA'}(X,\phi)$ in a similar
manner
to the construction of the general models [\HPT,\PT]. The action for the
theory is given by
$$\eqalign{
S = \int\! d^2x & \left\{ \eab\eyz \dmm X^{AY} \dpp
X^{BZ} +i\eabp\eyz \psi_-^{A'Y} \dpp \psi_-^{B'Z}
\right. \cr & \left.
+\eabp\eyzp\dmm \phi^{A'Y'} \dpp \phi^{B'Z'}
+i\eab\eyzp\chi_-^{AY'} \dpp \chi_-^{BZ'}  \right. \cr
& \left.
+i\lambda_+^a \dmm \lambda^a_+
-{im\over2}\lambda_+^a
\left(\epsilon^{BD}{\partial C^a_{BB'}\over \partial X^{DY}}\psi_-^{B'Y}
+\epsilon^{B'D'}{\partial C^a_{BB'}\over \partial
\phi^{D'Y'}}\chi_-^{BY'}\right)
 \right. \cr
& \left.
-{m^2\over8}\epsilon^{AB}\epsilon^{A'B'}C^a_{AA'}C^a_{BB'} \right\}\ , \cr}
\eqn\action
$$
where
$$
\dpp = {1\over{\sqrt{2}}}(\partial_0 + \partial_1 )\ \ \  \ \ \ \ \ \ \
\dmm = {1\over{\sqrt{2}}}(\partial_0 - \partial_1 )\ ,
$$
and $m$ is an arbitrary mass parameter. Note the
twisted form of the Yukawa interactions in \action\  in comparison to the
models
of [\HPT,\PT]. The free field theory ($m$ =0) possesses an
$SU(2)\times Sp(k)\times SU(2) \times Sp(k')$ rigid symmetry acting on the $AB,
YZ,A'B', Y'Z'$ indices respectively which is generally broken by the potential
terms.

Provided that $C^a_{AA'}$ takes the simple form
($M^a_{AA'}$,$N^a_{A'Y}$,$D^a_{AY'}$ and $E^a_{YY'}$ are constant tensors)
$$
C^a_{AA'} = M^a_{AA'} + \epsilon_{AB}N^a_{A'Y} X^{BY} +
\epsilon_{A'B'}D^a_{AY'}\phi^{B'Y'}  + \epsilon_{AB}\epsilon_{A'B'}E^a_{YY'}
X^{BY} \phi^{B'Y'} \ ,
\eqn\onesusy
$$
subject to the constraint
$$
C^a_{AA'}C^a_{BB'}+C^a_{BA'}C^a_{AB'} = 0 \ ,
\eqn\foursusy
$$
then the action \action\ has the on-shell (0,4) supersymmetry
$$\eqalign{
\delta X^{AY} &= i\epsilon_{A'B'}\eta^{AA'}_+\psi_-^{B'Y}\cr
\delta \psi^{A'Y}_- &= \epsilon_{AB}\eta^{AA'}_+\dmm X^{BY}\cr
\delta \phi^{A'Y'} &= i\epsilon_{AB}\eta^{AA'}_+\chi_-^{BY'}\cr
\delta \chi^{AY'}_- &= \epsilon_{A'B'}\eta^{AA'}_+\dmm \phi^{B'Y'} \cr
\delta \lambda^{a}_+ &= \eta^{AA'}_+C^a_{AA'} \ ,\cr}
\eqn\susy
$$
where $\eta^{AA'}_+$ is an infinitesimal anticommuting spinor
parameter. As is discussed by Witten [\Wittena], the above construction of
models
with (0,4) supersymmetry can be interpreted as a string theory analogue of the
ADHM construction of instantons with instanton number $k'$ in a target space
dimension of $4k$.

The general form of massive $(p,q)$ supersymmetric sigma models has been
discussed in terms of $(0,1)$ superfields in [\HPT,\PT] and we now provide such
a formulation of the ADHM model. To this end we introduce a tensor $I^A_{\
A'}$ satisfying
$$
\epsilon_{AB}I^A_{\ A'}I^B_{\ B'} = \epsilon_{A'B'}
\eqn\Idef
$$
which can be interpreted as a complex structure in the sense that
$I^{AB'}I_{AC'}=-\delta^{B'}_{C'}$, $I^{BA'}I_{CA'}=-\delta^{B}_{C}$.
The 'twisted' superfields are
$$\eqalign{
\cX ^{AY} &= X^{AY} + \theta^-I^A_{\ A'}\psi_-^{A'Y} \cr
\Phi^{A'Y'} &= \phi^{A'Y'} + \theta^-I^{\ A'}_{A}\chi_-^{AY'}\ \cr
\Lambda^a_+ &= \lambda^a_+ + \theta^-F^a \ ,\cr}
\eqn\superfields
$$
where $\theta^-$ is the anticommuting spinorial (0,1) superspace coordinate
with the associated superspace covariant derivative
$$
D_- = {\partial\over\partial \theta^-} + i\theta^-\dmm
$$
and $F^a$ is an auxiliary field. After removing $F^a$ by it's equation of
motion and using the constraint \foursusy, the action \action\ can be seen to
have the superspace form
$$\eqalign{
S_{effective} = -i\int\! d^2xd \theta^- & \left\{
 \eab\eyz D_-\cX^{AY}\dpp
\cX^{BZ} +\eabp\eyzp D_-\Phi^{A'Y'}\dpp \Phi^{B'Z'}
\right. \cr
&\left.
- i \delta_{ab}\Lambda^a_+ D_- \Lambda^b_+ - mC_a\Lambda^a_+
\right\} \ , \cr}
\eqn\suspaction
$$
where $C^a = I^{AA'}C^a_{AA'}$. Since the vector field $C^a$ is harmonic
and the target space is ${\bf R}^{4(k+k')}$, the model satisfies the
requirements of $(0,4)$ supersymmetry found in [\PT]. The inclusion of the
auxiliary field allows one to close a $(0,1)$ part of the supersymmetry
algebra off-shell. As with the component field formulation \action\ the
full $(0,4)$ supersymmetry is only on-shell. A manifestly off-shell form
requires harmonic superfields with an infinite number of auxiliary fields
[\GS].

Lastly we outline the $k=k'=1,\ n=8$ case (ie. a single instanton in ${\bf
R}^4$) analyzed by Witten which will be of primary interest here. The right
handed fermions are taken to be $\lambda^a_+ =
(\lambda^{AY'}_+,\lambda^{YY'}_+)$ and the tensor $C^a_{AA'}$ takes the form
$$\eqalign{
C^{YY'}_{BB'} &= \epsilon_{BC}\epsilon_{B'C'}X^{CY}\phi^{C'Y'} \cr
C^{AY'}_{BB'} &= {\rho\over\sqrt2}\epsilon_{B'C'}\delta^A_B\phi^{C'Y'}\ ,
\cr}
$$
where $\rho$ is an arbitrary constant to be interpreted as the instanton size.
The bosonic potential for this theory is easily worked out as
$$
V = {m^2\over8}(\rho^2 + X^2)\phi^2 \ ,
\eqn\V
$$
where $X^2 = \epsilon_{AB}\epsilon_{YZ}X^{AY}X^{BZ}$ and similarly for
$\phi^2$. Thus, for $\rho \ne 0$, the vacuum states of the theory are
defined by $\phi^{A'Y'}=0$, and parameterize ${\bf R}^4$. The
$X^{AY}$ and $\psi_-^{A'Y}$ are massless fields while $\phi^{A'Y'}$ and
$\chi_-^{AY'}$ are massive. This yields exactly 4 of the
$\lambda^a_+$ massive and 4 massless.

\chapter{Quantization}

\subsection{Renormalization}

It is not hard to see that the model described above is superrenormalizable
in two dimensions as the interaction vertices do not carry any momentum
factors and have at most three legs. In fact a little inspection reveals that
the only possible divergences of the theory are the one loop graphs
contributing to the potential. This can also been seen from the following
simple superspace argument, valid for any $(0,1)$ supersymmetric
linear massive sigma model. The superspace measure $d^2xd\theta^-$ has mass
dimension $-3/2$ while all vertices contribute a factor of $m$ to the
effective action. Thus by power counting, only graphs with a single vertex
can yield divergent contributions to the effective action. Of these, only the
one loop (tadpole) graphs are relevant, all the higher loop divergences are
removed by the renormalization procedure.

Although the potential provides masses for some
of the fields there will in general be massless fields which may cause
infrared divergences. We must therefore add an infrared regulator in the
form of a mass $M$ to the propagator and treat any mass terms in \action\ as
interactions, taking the limit $M\rightarrow 0$ in the final expressions.
Using dimensional regularization in $D=2+\epsilon$ dimensions and the
background field method [\Back] the bosonic graphs are readily calculated to
be
$$\eqalign{ \Gamma_{Div}({\rm bosons})=-&{m^2\over4}\Delta(0)
\left[\epsilon^{AB}\epsilon^{CD}\epsilon^{C'D'}\epsilon^{YZ} {\partial
C^a_{CC'}\over\partial X^{AY}}  {\partial C^a_{DD'}\over\partial X^{BZ}}
\right.\cr  & \left. \ \ \ \ \ \
+\epsilon^{A'B'}\epsilon^{C'D'}\epsilon^{CD}\epsilon^{Y'Z'}  {\partial
C^a_{CC'}\over\partial \phi^{A'Y'}}  {\partial C^a_{DD'}\over\partial
\phi^{B'Z'}} \right] \ , \cr}   \eqn\bosons
$$
while the fermionic graphs are
$$\eqalign{
\Gamma_{Div}({\rm fermions}) =&{m^2\over4}\Delta(0)
\left[{1\over2}\epsilon^{AC}\epsilon^{BD}\epsilon^{C'D'}\epsilon^{YZ}
{\partial C^a_{CC'}\over\partial X^{AY}}
{\partial C^a_{DD'}\over\partial X^{BZ}} \right. \cr
& \left. \ \ \ \ \ \
+{1\over2}\epsilon^{A'C'}\epsilon^{B'D'}\epsilon^{CD}\epsilon^{Y'Z'}
{\partial C^a_{CC'}\over\partial \phi^{A'Y'}}
{\partial C^a_{DD'}\over\partial \phi^{B'Z'}} \right]\ , \cr}
\eqn\fermions
$$
where all the tensor expressions are evaluated at the background fields and
the bosonic propagator at zero momentum and renormalization scale $\mu$ is
$$
\Delta (0) = -{1\over 2\pi\epsilon}-{\rm ln}\left({M^2\over\mu^2}\right)
+{\rm finite} \ .
$$

One can see that the epsilon tensor terms in \bosons\ and \fermions\ are
different as a result of the twisted form of the Yukawa interactions. It is
not immediately obvious then that the bosonic and fermionic divergences
cancel. However by substituting in \onesusy\ it is not much trouble to see
that they do and hence $\Gamma_{Div}=0$. Thus Witten's ADHM model is
ultraviolet finite to all orders of perturbation theory. Therefore there is
no renormalization group flow in these models. This result may be expected,
but is not guaranteed, by supersymmetry as there is a general argument for
finiteness only for off-shell $(0,4)$ sigma models, with some modifications
required due to anomalies [\HPb].

\subsection{Integrating the Massive Modes}

In this section we will integrate out the massive modes. We shall postpone
the problem of anomalies in chiral supersymmetric models until
the next section. We assume for simplicity here that
$M^a_{AA'}=N^a_{AY}=0$, $D^a_{A'Y'} \ne 0$ so that the $X^{AY}$ and
$\psi^{A'Y}_-$ fields are massless, the $\phi^{A'Y'}$ and $\chi_-^{AY'}$
fields massive and the vacuum is at $\phi^{A'Y'}=0$. The theory is then only
quadratic in the massive fields and integrating over them is therefore exact
at the one loop level. This assumption also ensures that the interacting
theory breaks the $SU(2) \times Sp(k) \times SU(2) \times Sp(k')$ symmetry of
the free theory down to $SU(2) \times Sp(k')$ [\Wittena]. We may therefore
write
$$\eqalign{
C^a_{AA'} &=\epsilon_{A'B'}D^a_{AY'}\phi^{B'Y'}
+ \epsilon_{AB}\epsilon_{A'B'}E^a_{YY'} X^{BY}\phi^{B'Y'} \cr
&\equiv \epsilon_{A'B'}B^a_{AY'}(X)\ \phi^{B'Y'} \ . \cr}
$$
At this point it is necessary to split up the left handed fermions into
there massive and massless parts. If we introduce the zero modes $v^a_i(X)$,
$i=1,2,...,n-4k'$ of the fermion mass matrix, defined such that
$$
v^a_iB^a_{AY'} = 0,   \ \ \ \ v^a_iv^a_j = \delta_{ij}
$$
and a similar set of massive modes $u^a_I(X)$, $I=1,2,...,4k'$ satisfying
$$
u^a_Iu^a_J = \delta_{IJ}, \ \ \ \ u^a_Iv^a_j =0
$$
then we may separate the $\lambda^a_+$ as
$$
\lambda^a_+ = v^a_i\zeta^i_+ + u^a_I\zeta^I_+ \ .
\eqn\lamodes
$$
so that the $\zeta^i_+$ are massless and the $\zeta^I_+$ massive.
We now rewrite the action \action\ in terms of the massless and massive
fields
$$
S = S_{massless} + S_{massive}
\eqn\S
$$
where $S_{massless}$ is the part of \action\ which only contains the
massless fields. Explicitly
$$
\eqalign{
S_{massless} = \int\! d^2x & \left\{  \eab \eyz \dmm X^{AY} \dpp X^{BZ}
+ i \eabp \eyz \psi^{A'Y}_- \dpp \psi^{B'Z}_-
\right. \cr
&\left.
+i\zeta^i_+ (\delta_{ij}\dmm\zeta^j_+ + A_{ijAY}\dmm X^{AY}\zeta^j_+)
\right\} \ ,\cr}
\eqn\Smassless
$$
where
$$
A_{ijAY}=v^a_i{\partial v^a_j\over\partial X^{AY}}
\eqn\Adef
$$
is the induced $SO(n-4k')$ connection and
$$\eqalign{
S_{massive} = \int\! & d^2x \left\{ \eabp \eyzp
\dmm\phi^{A'Y'}\dpp\phi^{B'Z'} + i\eab \eyzp \chi^{AY'}_- \dpp \chi^{BZ'}_-
\right. \cr
& \left.
+i\delta_{IJ}\zeta^I_+\dmm \zeta^J_+ + iA_{IJAY} \dmm X^{AY}
\zeta^I_+ \zeta^J_+
+2iA_{iJAY} \dmm X^{AY}
\zeta^i_+\zeta^J_+ \right. \cr
& \left.
-im\epsilon_{B'C'}v^a_iE^a_{YY'}\zeta^i_+\phi^{C'Y'}\psi^{B'Y}_-
-im\epsilon_{B'C'}u^a_IE^a_{YY'}\zeta^I_+\phi^{C'Y'}\psi^{B'Y}_-
\right. \cr
& \left.
-imu^a_{I}B^a_{BY'}\zeta^I_+\chi_-^{BY'}
-{m^2\over8}\epsilon^{AB}\epsilon_{C'D'}
B^a_{AY'}B^a_{BZ'}\phi^{C'Y'}\phi^{D'Z'} \right\} \ ,\cr}
\eqn\Smassive
$$
where $A_{IJAY}=u^a_I{\partial u^a_J}/ {\partial X^{AY}}$ and
$A_{iJAY}=v^a_i{\partial u^a_J}/ {\partial X^{AY}}$.

The classical low energy effective action is simply obtained by considering
the most general action possible which is compatible with all of the
symmetries of the theory. To calculate the effective action quantum
mechanically we will integrate over the massive fields and discard any higher
derivative terms. The presence of higher derivative terms in the effective
action, which are suppressed by powers $p/m$ where $p$ is the low energy
momentum scale, would ruin the renormalizability and prevent a simple
geometrical sigma model interpretation of the effective theory.

First we notice that because of the nontrivial definition of the massless
left handed fermions \lamodes , $S_{massless}$ is not (0,4) supersymmetric by
itself as it is missing a four fermion interaction term. The problem is
rectified by noting that there is a tree graph, with a single internal
$\phi^{A'Y'}$ field propagating, which contributes to the low energy effective
action. In order to avoid the singular behaviour of the propagator at zero
momentum, when calculating this graph it is helpful to use the massive
propagator for $\phi^{A'Y'}$, obtained from the last term in \Smassive .

At this point it is useful to write, using \foursusy\ ,
$$
B^a_{AY'}B^a_{BZ'} = \epsilon_{AB}\epsilon_{Y'Z'}\Omega
$$
where
$$
\Omega(X)={1\over4k'}\epsilon^{AB}\epsilon^{Y'Z'}B^a_{AY'}B^a_{BZ'} \ .
\eqn\Om
$$
The last term in \Smassive\ becomes
$$
-{m^2\over4}\Omega(X)\ \phi^2 \ ,
$$
and hence can be interpreted as the $X^{AY}$
dependent mass term for $\phi^{A'Y'}$. The tree graph can then be seen to
contribute the four fermion term
$$
-{1\over2}\zeta^i_+ \zeta^j_+F^{ij}_{A'YB'Z}\psi^{A'Y}_- \psi^{B'Z}_-
$$
where
$$
F^{ij}_{A'YB'Z} =
2 \epsilon_{A'B'} \epsilon^{Y'Z'}\Omega^{-1}
v^a_i E^a_{(Y|Y'} v^b_j E^b_{|Z)Z'}
 \ ,
\eqn\Fdef
$$
which we will later relate to the field strength tensor.

We may now discard all vertices with only one massive field in \Smassive\ and
examine the one loop contributions to the effective action. Inspection of the
quadratic terms in  $S_{massive}$ shows there are no contributions to the
gauge connection in \Smassless. Furthermore, inspection shows that of all
the other possible contributions only those corresponding to the effective
potential do not involve higher order derivatives of the massless fields.
A check on this is to note that any terms which are second
order in the derivatives are logarithmically divergent, and by finiteness
of the model, must therefore vanish.

To calculate the effective potential we simply
set $\dmm X^{AY} = \dpp X^{AY}= \psi^{A'Y}_-=0$. Thus only the last two terms
in  \Smassive\ need be considered (we no longer use a massive propagator for
$\phi^{A'Y'}$). The effective potential then receives the standard bosonic
and fermionic contributions (in Euclidean momentum space)
$$
V_{eff}({\rm bosons}) = {\alpha'\over4\pi}\sum_{n=1}^{\infty}{1\over n} \int\!
d^2 p\  {\rm Tr}\left[{\epsilon_{C'D'} \epsilon^{AB}B^a_{AY'}B^a_{BZ'}\over
4p^2/m^2}\right]^n  \eqn\Veffbose
$$
and
$$
V_{eff}({\rm fermions}) = -{\alpha'\over4\pi} \sum_{n=1}^{\infty}{1\over n}
\int\! d^2 p\ {\rm Tr}\left[{ u^a_Iu^{bI}B^a_{CY'}B^b_{DZ'}\over 2 p^2/m^2}
\right]^n \ .
\eqn\Vefffermi
$$
Now the definition \Om\ yields the following expressions
$$
\epsilon_{C'D'}\epsilon^{AB}B^a_{AY'}B^a_{BZ'} =
2\epsilon_{C'D'}\epsilon_{Y'Z'}\Omega
$$
and
$$
u^a_Iu^{bI}B^a_{CY'}B^b_{DZ'} =
\epsilon_{CD}\epsilon_{Y'Z'}\Omega \ .
$$
Therefore \Veffbose\  is completely canceled by \Vefffermi\ and there is no
contribution to the effective potential.

{}From the above analysis we conclude that the effective quantum action of the
massless fields is
$$\eqalign{
S_{effective} & = \int\! d^2x \left\{ \eab \eyz \dmm X^{AY} \dpp X^{BZ}
+ i\eabp \eyz \psi^{A'Y}_- \dpp \psi^{B'Z}_-
\right. \cr
&\left. \ \ \
+i\zeta^i_+ (\delta_{ij}\dmm \zeta^j_+ + A_{ijAY}\dmm X^{AY}\zeta^j_+)
- {1\over2}\zeta^i_+ \zeta^j_+F^{ij}_{A'YB'Z}\psi^{A'Y}_- \psi^{B'Z}_-
\right\} \ .\cr}
\eqn\effective
$$
This is simply the action of the general (0,4) supersymmetric nonlinear sigma
model [\HPT,\PT], although the right handed superpartners of $X^{AY}$ are
'twisted'. As with the original theory \action , the low energy effective
theory \effective\ admits a $(0,1)$ superfield form. Introducing the
superfield $\Lambda^i_+ =\zeta^i_+ + \theta^-F^i$ then allows us (after
removing $F^i$ by it's equation of motion) to express \effective\ as
$$\eqalign{
S_{effective} = -i\int\! d^2xd & \theta^- \left\{
\eab\eyz D_-\cX^{AY}\dpp \cX^{BZ} \right. \cr
& \left.\ \ \ \ \
- i\Lambda^i_+ (\delta_{ij} D_- \Lambda^j_+
+ A_{ijAY} D_-\cX^{AY}\Lambda^j_+) \right\} \ , \cr}
\eqn\superaction
$$
provided that $F^{ij}_{A'YB'Z}$ satisfies
$$
F^{ij}_{A'YB'Z} = I^A_{\ A'}I^B_{\ B'}F^{ij}_{AYBZ}
\eqn\constraint
$$
where $F^{ij}_{AYBZ}$ is the curvature of the connection \Adef ,
$$
F^{ij}_{AYBZ} = \partial_{AY} A_{ijBZ} - \partial_{BZ} A_{ijAY}
+ A_{ikAY}A_{kjBZ}- A_{ikBZ} A_{kjAY}\ .
$$
This is just the familiar constraint on (0,4) models that the field strength
be compatible with the complex structure [\HPT,\PT]. Furthermore it is not hard
to check that $S_{effective}$ does indeed possess the
full on-shell (0,4) supersymmetry (the superspace formulation \superaction\
ensures only off-shell (0,1) supersymmetry) precisely when \constraint\ is
satisfied.

For the $k=k'=1,\ n=8$ model above it is straightforward to determine
the non zero components of $v^a_i$ and $u^a_I$ as
$$
\eqalign{
v^{YY'}_{ZZ'}& =
{\rho\over\sqrt{\rho^2+X^2}}\delta^Y_Z \delta^{Y'}_{Z'}\ \ \ \ \ \
v^{AY'}_{ZZ'} =
-{\sqrt{2}\over\sqrt{\rho^2+X^2}}X^{A}_{\ Z}\delta^{Y'}_{Z'}
\cr
u^{YY'}_{BZ'}& = {\sqrt{2}\over\sqrt{\rho^2+X^2}}X_{B}^{\ Y}\delta^{Y'}_{Z'}
\ \ \ \ \ \
u^{AY'}_{BZ'} = {\rho\over\sqrt{\rho^2+X^2}}\delta^A_{B}\delta^{Y'}_{Z'}
\ ,\cr}
\eqn\uv
$$
and the mass term \Om\ is
$$
\Omega ={1\over2}(X^2 + \rho^2) \ .
$$
The gauge field $A_{ijAY}$ obtained from \uv\ is simply that of
a single instanton on the manifold ${\bf R}^4$
$$
A^{YY'ZZ'}_{AX} = -
\epsilon^{Y'Z'}
{(\delta^{Z}_{X}X_{A}^{\ \ Y}
+ \delta^{Y}_{X}X_{A}^{\ \ Z})\over\rho^2+X^2} \ ,
\eqn\A
$$
and the four fermion vertex \Fdef\ is
$$
F^{TT'UU'}_{A'YB'Z} = {4\rho^2\over(X^2 + \rho^2)^2}
\eabp \epsilon^{T'U'} \delta^{T}_{(Y}\delta^{U}_{Z)}\ ,
\eqn\F
$$
which is precisely the field strength of an instanton, justifying our
presumptuous notation, and can be easily seen to satisfy \constraint .

\subsection{Anomalies}

So far we have ignored the possibility of anomalies in the quantum theory.
While the original theory \action\ is simply a linear sigma model and
therefore possesses no anomalies, this is not the case for the effective theory
\effective. It is well known that off-shell $(0,4)$ supersymmetric sigma models
suffer from chiral anomalies which break spacetime gauge and coordinate
invariance, unless the gauge field can be embedded in the spin connection of
the target space. In addition, working in $(0,1)$ superspace
only ensures that $(0,1)$ supersymmetry is preserved and there are
also extended supersymmetry anomalies as the $(0,4)$ supersymmetry is not
preserved. We therefore expect that we will have
to add finite local counter terms to \effective\  at all orders of perturbation
theory so as to cancel these anomalies. This requires that the spacetime metric
and antisymmetric tensor fields become non trivial at higher orders of
$\alpha'$, while on the other hand the gauge connection is unaffected [\HPa].

An alternative way of viewing this is to note that although the action
\effective\ is classically conformally invariant, when quantized it may not be
ultraviolet finite and hence break scale invariance. There is a power counting
argument which asserts that off-shell $(0,4)$ supersymmetric models are
ultraviolet finite to all orders of perturbation theory [\HPb]. This
argument is further complicated by sigma model anomalies and it has been stated
that only the non chiral models are ultraviolet finite. Indeed while
off-shell $(0,4)$ supersymmetric theories are one loop finite,
there is a two loop contribution of the form ${\rm Tr}(R^2 -F^2)$ [\NDL] which
certainly does not vanish in general. Normally this leads
to non vanishing $\beta$-functions which we must then take into account when
determining the conformal fixed point of the renormalization group flow.
However, in models with off-shell $(0,4)$ supersymmetry the non vanishing
$\beta$-functions can be canceled by redefining the spacetime fields at each
order of $\alpha'$, in such a way as to ensure that supersymmetry is
preserved in perturbation theory [\HPa]. This has been well studied and
verified up to three loops. Thus the ultraviolet divergences which arise in
the quantization of off-shell $(0,4)$ models are really an artifact of the
use of a renormalization scheme which does not preserve the supersymmetry.
The off-shell $(0,4)$ models are ultraviolet finite in an appropriate
renormalization scheme.

However the model here has only on-shell $(0,4)$ supersymmetry and these
finiteness arguments do not immediately apply. At least in the $k=k'=1$
case however the gauge group $SO(4)\cong SU(2)\times SU(2)$ contains a subgroup
$Sp(1)\cong SU(2)$ which admits three complex structures obeying the
algebra of the quaternions. This endows the target space of the left handed
fermions with a hyper Kahler structure and facilitates an off-shell
formulation using constrained superfields [\HPc]. We may therefore expect
that it is ultraviolet finite in the same manner as the off-shell models
described above.

In [\HPa] the necessary field redefinitions were derived to order $\a^2$ for
$(0,4)$ supersymmetric sigma models. Both the target space metric and
antisymmetric tensor field strength receive corrections to all orders in
$\a$. Howe and Papadopoulos found that in order to maintain $(0,4)$
supersymmetry in perturbation theory the target space metric (which is flat
here at the classical level) must receive corrections in the form of a
conformal factor
$$
\eab \eyz \rightarrow
\left(1 - {3\over2}\a f - {3\over16}\a^2 \triangle f + ... \right)
\eab \eyz\ .
\eqn\conf
$$
They also showed, up to three loop order, that these redefinitions cancel the
ultraviolet divergences which arise when one renormalizes \effective\
using standard $(0,1)$ superspace methods, which do not ensure
(0,4) supersymmetry is preserved perturbatively. In addition, the
antisymmetric field strength tensor becomes $H = -{3\over4}\a*df$ so as to
cancel the gauge anomaly
$dH=-{3\a\over4}{\rm Tr} F \wedge F$. Furthermore there are no
corrections to the instanton gauge field.

For the instanton number one model considered here, Howe and Papadopoulos give
the
function $f$ as
$$
f = -\triangle {\rm ln}(X^2 + \rho^2)\ ,
$$
where $\triangle$ is the flat space Laplacian. It is a simple matter to
calculate the conformal factor \conf\ and hence the  target space metric as
$$
g_{AYBZ} = \left(1 + 6\a{X^2+2\rho^2 \over(X^2+\rho^2)^2}
- 18\a^2 {\rho^4 \over(X^2+\rho^2)^4}+...
\right)\eab \eyz \ . \eqn\metric
$$
To order $\a$ this is the solution of Callan, Harvey and Strominger
[\CHS] obtained by solving the first order equations of motion of the
10 dimensional heterotic string (although with $n=6$ rather than $n=8$
in their notation). Thus the target space has been curved around the instanton
by
stringy effects but remains non singular so long as $\rho \ne 0$. The case
$\rho=0$ is of great interest as it provides a string theoretic
compactification of instanton moduli space. We will briefly discuss this in
the next section.

\chapter{Concluding Remarks}

In the above we found the order $\a^2$ corrections to the low energy
effective action of the ADHM sigma model. Such solutions have been discussed
before [\CHS] and we agree with their solution to first order. In our
calculations we have expanded in the parameter
$$
\a\Omega^{-1}={2\a\over{X^2 +\rho^2}}
$$
and hence our approximations are valid for all $X$ if $\rho^2 \gg \a$ and for
$X^2 \gg \a$ even when $\rho^2$ is small. An interesting question
raised is what are the stringy corrections to the classical instanton in the
extreme case that it's size vanishes? One can see from \metric\ that the
order $\a$ corrections persist when $\rho=0$ so the effective theory is non
trivial. It has been conjectured that there should be a $(4,4)$
supersymmetric sigma model for instantons of zero size [\GS,\GSt] which could
be constructed from a massive linear $(4,4)$ supersymmetric model. In [\GSt]
the conditions for the ADHM model to possess full (4,4) supersymmetry in the
infrared limit were derived. There it was found that the metric must be
conformally flat, with the metric satisfying Laplace's equation. This is in
agreement with what we have found here in the $\rho=0$ case above (see \conf\
and \metric ) and lends some additional support to this conjecture.

If we start with the linear sigma model \action\ with $\rho=0$ the
$\lambda_+^{AY}$ fields are massless and decouple from the theory. The vacuum
states are defined by $X^{AY}=0$ or $\phi^{A'Y'}=0$ and there is a symmetry
between $X^{AY}$ and $\phi^{A'Y'}$. Let us assume we choose the
$\phi^{A'Y'}=0$ vacuum. Then as before the fields $X^{AY}$ and $\psi_-^{AY'}$
are massless and the $\phi^{A'Y'}$, $\chi_-^{AY'}$ and $\lambda_+^{YY'}$
fields all have masses $ m\sqrt{X^2/2}$. Upon integrating out the massive
fields we would simply obtain a free field theory, which trivially
possesses $(4,4)$ supersymmetry. At the degenerate vacuum $X^{AY}=0$
however, all fields are massless and there is a single interaction term
$m\lambda_{+YY'}\phi^{\ \  Y'}_{A'}\psi_-^{A'Y}$. Thus the moduli space of
vacua does not have a manifold structure. For $X^{AY} \ne 0$ the vacuum
states are simply ${\bf R}^4$ but at the point $X^{AY} = 0$ lies another
entire copy of ${\bf R}^4$ (parameterized by the $\phi^{A'Y'}$). This odd
state of affairs is smoothly resolved if we first construct the effective
theory and then take the limit of vanishing instanton size.

Let us now take the limit $\rho \rightarrow 0$ of the effective action
\effective . It should be noted that the Yang-Mills instanton has shrunk to
zero size but it has not disappeared in the sense that the topological charge
remains equal to one. Unfortunately our expressions are not a priori valid
near $X=0$. Nevertheless we will try to shed some light about what the
complete string theory solution could be in that region. When $\rho$ vanishes
both the field strength \Fdef\ and the $O(\a)$ sigma model anomaly vanish. We
are however, still left with a non trivial metric and anti symmetric tensor.
It seems reasonable to assume then that all the anomalies are canceled by
these. The metric then has the exact conformal factor
$$
g_{\mu\nu} = \left(1 - {3\a\over2}f\right)\delta_{\mu\nu} \ ,
\eqn\cfact
$$
and antisymmetric field
$$
H_{\mu\nu\rho} = -{3\a\over4}\epsilon_{\mu\nu\rho\lambda}\partial^{\lambda}f \
,
\eqn\H
$$
where $f = -4/X^2$ and we have switched to a more convenient notation. This
geometry is similar to the one discussed by Callan, Harvey and
Strominger [\CHS], although the anti symmetric field is not the same
and leads to a different interpretation in the limit $X^2 \rightarrow 0$ as we
will shortly see. The target space is non singular, asymptotically flat and has
a
semi-infinite tube with asymptotic radius $\sqrt{6\a}$, centered around the
instanton. That is to say the apparent singularity at $X^2=0$ in \metric\ is
pushed off to an "internal infinity" down the infinite tube. Thus the
problematic
$X^{AY}=0$ vacua are pushed an infinite distance away and the manifold
structure
is preserved. The resolution of this description with the non manifold picture
described above has been discussed by Witten [\Wittenb].

In the limit $X^2 \ll \a$ the modified spin connection with torsion becomes,
where $\alpha,\beta$ are vierbein indices,
$$
\eqalign{
\omega_{\mu}^{(-)\alpha\beta} &\equiv \omega_{\mu}^{\alpha\beta}
+ H_{\mu}^{\ \alpha\beta} \cr
&=
-(\delta^\alpha_{\mu} \delta^\beta_{\nu}
-\delta^\alpha_{\nu} \delta^\beta_{\mu}
+ \epsilon_{\mu\nu}^{\ \ \ \alpha\beta}){X^{\nu}\over X^2}
\cr } \ ,
\eqn\ominus
$$
which is a flat connection! That is to say far down the infinite tube
the torsion parallelizes the manifold (which is asymptotically $S^3 \times {\bf
R}$ and is indeed parallelizable). The gauge connection is also flat (for
$X^{AY}
\ne 0$) and can therefore be embedded into the generalized spin connection
\ominus\ (they are both $so(4)$ valued). The low energy effective
theory therefore possesses $(4,4)$ supersymmetry in the region $X^2
\rightarrow 0$ and is free of anomalies there. This supports our assumption
that the anomalies are canceled and the expressions \cfact\ and \H\ are
exact, at least in this region. For the region $X^2 \gg \a$ our perturbative
expansion is valid and the the theory only possesses $(0,4)$ supersymmetry
since the gauge connection can not be embedded into the spin connection.
Although in a similar spirit in the limit $X^2 \rightarrow \infty$ the
curvatures vanish and the theory is free and again has $(4,4)$
supersymmetry. In a sense then the $\rho=0$ ADHM instanton can be viewed as
a soliton in the space of string vacua interpolating between two $(4,4)$
supersymmetric sigma models, just as the target space can be viewed as
interpolating between two supersymmetric ground states of supergravity
[\GT].

However, since we still have instanton
number one, the vector bundle is non trivial whereas the tangent bundle is
trivial. Thus in the region $X^2 \rightarrow 0$ we can only identify the spin
connection with the gauge connection locally. $(4,4)$ supersymmetry is then
broken by global, non perturbative effects back to $(0,4)$ supersymmetry.
This is reflected by the observation [\GT] that the $S^3 \times M^7$
compactification of $D=10$ supergravity breaks half the supersymmetry.

I would like to thank G. Papadopoulos and P.K. Townsend for their advice and
Trinity College Cambridge for financial support.

\refout

\end